\documentclass[onecolumn,showpacs,preprintnumbers,amsmath,amssymb]{revtex4}

\def\bi{\begin{itemize}}
\def\ei{\end{itemize}}
\def\bq{\begin{quotation}}
\def\eq{\end{quotation}}

\def\thedemobiblio#1{\smallskip\par
 \list{}{\labelwidth 0pt \leftmargin 1em \itemindent -1em \itemsep 1pt}
 \small \parindent 0pt
 \parskip 1.5pt plus .1pt\relax
 \def\newblock{\hskip .11em plus .33em minus .07em}
 \sloppy\clubpenalty4000\widowpenalty4000
 \sfcode`\.=1000\relax}

\usepackage{psfrag}
\usepackage{graphicx}% Include figure files
\usepackage{dcolumn}% Align table columns on decimal point
\usepackage{bm}% bold math
\newcommand{\bfg}[1]{\mbox{\boldmath $#1$}}
\newcommand{\union}{\mathop{\bigcup}\limits}

\begin{document}

\title{A Concurrent Multiscale Micromorphic Molecular Dynamics.\\ Part I. Theoretical Formulation}

\author{Shaofan Li}
\email{shaofan@berkeley.edu}

\author{Qi Tong}

\affiliation{Department of Civil and Environmental Engineering,\\
University of California, Berkeley, CA94720, USA }

\date{March 10, 2014}

\begin{abstract}
Based on a novel concept of multiplicative multiscale
decomposition, we have derived
a multiscale micromorphic molecular dynamics (MMMD)
to extent the (Andersen)-Parrinello-Rahman molecular dynamics
to mesoscale and macroscale.
The multiscale micromorphic molecular dynamics
is a con-current three-scale particle dynamics that couples
a fine scale molecular dynamics,
a mesoscale particle dynamics of micromorphic medium, and
a coarse scale nonlocal particle dynamics of nonlinear continuum 
together.
By choosing proper statistical closure conditions,
we have shown that
the original Andersen-Parrinello-Rahman molecular dynamics
can be rigorously formulated and justified from first principle,
and it is a special case of the proposed multiscale micromorphic molecular dynamics.
The discovered mutiscale structure and the corresponding multiscale dynamics
reveal a seamless transition from atomistic scale to continuum scale
and the intrinsic coupling relation among them.
The proposed MMMD can be used to solve
finite size nanoscale science and engineering problems with arbitrary boundary
conditions.
\end{abstract}


\maketitle

\section{Introduction}
Thirty some years ago, in his seminal work Andersen \cite{Andersen80}
first proposed an isoenthalpic-isobaric ensemble of
molecular dynamics (MD) allowing
the volume a cubic lattice cell to vary.
Subsequently, Parrinello and Rahman \cite{PR80,PR81}
elegantly extended Andersen's
formalism to anisotropic cases allowing both the volume and the shape of
a molecular dynamics (MD) cell to vary.
Since then the (Andersen)-Parrinello-Rahman (APR) molecular
dynamics has become the standard molecular dynamics method
in simulations of structural transformation and phase transitions.

However, the APR-MD approach has not been thoroughly understood, and
this is reflected in both its physical foundation as well as
how to extend it beyond the restriction of periodic boundary condition
so that it can bridge up to macroscale or continuum scale.
Moreover, the APR-MD lagrangian has been viewed as an ad hoc choice,
as Parrinello and Rahman commented in \cite{PR81},
``{\it $\cdots \cdots$
Whether such a Lagrangian is derivable from first principles is a question
for further study; its validity can be judged, as of now, by the equations
of motion and the statistical ensembles that it generates. $\cdots \cdots $
}''

In recent years, there have been renewed interests
in revising APR molecular dynamics,
e.g. \cite{Parrinello02,Parrinello03,Parrinello06,Podio08},
which attempted to extend APR MD to non-equilibrium condition or
to macroscale simulation.
On the other hand, there have been many efforts to formulate
multiscale coarse-grained molecular dynamics,
among them
\cite{Rudd-Broughton1,Rudd-Broughton2,Youping03,Youping09,Voth1,Voth2}
are probably most rigorous approaches.

One of the objectives of the present work is
to revise or generalize APR molecular dynamics so that it can
lead to a rigorous macroscale nonlinear continuum mechanics,
e.g. \cite{Podio08,DiCarlo09}. By doing so, one
hopes to achieve a multiscale molecular dynamics that
is not necessary to impose the periodic boundary condition.

In this paper, we revisit the topic again in an attempt to
accomplish the following two objectives:\\
(1) Derive the original (Andersen)-Parrinello-Rahman (APR)
Lagrangian from first principle,
and (2) Extend the single-cell APR molecular dynamics to a
multiscale continuum-molecular dynamics so that it is capable of simulating
lattice dynamic motions in multiple cells without imposing periodic boundary
condition, and bridging molecular dynamics
and nonlinear continuum mechanics from first principle.

The paper is arranged into seven Sections. In the second Section,
we discuss the multiscale decomposition that includes both micromorphic
decomposition (multiplicative decomposition) and the addictive decomposition.
In the third section, we discuss the random variables and their
 statistical closure. Section 4 restates the original PR-MD formulation.
In Section 5, we derive the general formulae for the multiscale micromorphic
molecular dynamics;
and in Section 6, we discuss a coarse-grain approach to
approximate the multiscale molecular dynamics when the fine scale
dynamics are turned off --- that is how
to obtain a rigorous coarse grained molecular dynamics.
Finally, in Section 7, we close the presentation by making
few comments and remarks.

\section{Multiscale decompositions}
The key to extend the APR molecular dynamics to continuum scale is
how to decompose the material displacement field into
different scales, and so that we can
establish the coupled dynamics equations
of displacement field at different scales.
In fact, the main reason why the PR-MD has been successful
is its ingenious multiscale decomposition of the displacement field
that reveals connections between mesoscale thermodynamics
with atomistic motion at the fine scale.

Before we construct any multiscale PR-MD, it would be
instrumental to discuss the multiscale decomposition
of the material displacement field first.
In the literature,
there are two types of multiscale decompositions in molecular dynamics
simulations:
(1) Micromorphic decomposition, or multiplicative decomposition,
and (2) Wagner-Liu's additive decomposition.

To start with, we first
assume that the crystalline material has periodic lattice
structure and its deformation is locally
uniform so that we can discretize the entire
domain into many cells, $\alpha = 1,2,\cdots, N_c$,
such that at macroscale the coarse continuum deformation in
each cell is spatially uniform.

We assume that for the i-th atom inside the $\alpha$-th cell,
its absolute displacement follows the following micromorphic
decomposition,
\[
{\bf r}_i (t)= {\bf r}_{\alpha} ({\bf R},t)
+ {\bf r}_{\alpha i} ({\bf R}_i, t),
~i=1,2,\cdots N_a
\]
where ${\bf r}_{\alpha}$ is the spatial position of the center of
mass in the $\alpha$-th cell, which may be identified as or
a function of coarse scale coordinate ${\bf R}$.
When $t=0$, we have
\begin{equation}
{\bf R}_i = {\bf R}_{\alpha} + {\bf R}_{\alpha i} ,
~i=1,2,\cdots N_a
\label{eq:decomposition-1}
\end{equation}
Suppose there are $N_{\alpha}$ atoms in the $\alpha$-th cell,
the center of mass in the $\alpha$-th cell
at the referential configuration
of the medium $\Omega_{\alpha}(0)$ is
\begin{equation}
{\bf R}_{\alpha} =
{1 \over \displaystyle \sum_{i} m_i } \sum_{i} m_i {\bf R}_{i}, ~~\forall
{\bf R}_{\alpha i} \in \Omega_{\alpha} (0)
\label{eq:centerM}
\end{equation}
We can then justify the decomposition Eq. (\ref{eq:decomposition-1}),
Note that Eq. (\ref{eq:centerM}) is not an additional constraint,
but an average property. That is the weighted
average micromorphic coordinates of a ensemble system is a single
coordinate field, i.e. coarse scale coordinate.
Whereas the average of the fine scale coordinates is zero,
\[
\displaystyle \sum_{i} m_i {\bf R}_{\alpha i} = 0.
\]
Similar decomposition is hold in the current configuration
$\Omega(t) = \union_{\alpha} \Omega_{\alpha}(t)$, i.e.
\begin{equation}
{\bf r}_i = {\bf r}_{\alpha} + {\bf r}_{\alpha i}, ~{\rm and}~
{\bf r}_{\alpha}= {1 \over \displaystyle \sum_{i} m_i}
\sum_{i} m_i {\bf r}_i
\end{equation}
By definition,
\begin{equation}
\sum_{i} m_i {\bf r}_{\alpha i} = 0,~~{\rm and}~~
\sum_{i} m_i \dot{\bf r}_{\alpha i} = 0~.
\end{equation}

Note that  ${\bf r}_{\alpha}$ is the position of center of mass
of the $\alpha$-cell, and it can be used to
represent the displacement field in the coarse scale
dynamics equations, which is one of the focus of this work.

We consider the so-called Micromorphic Decomposition.
In the micromorphic theory, a micro deformation tensor
is assigned to each cell, i.e. ${\bfg \phi}^{\alpha}=
\phi^{\alpha}_{ij} {\bf e}_i \otimes {\bf e}_j$,
which characterizes the microstructure of the cell.
In the PR-MD theory, the relative position of
each atom in a deformed cell is
is related to the micro-deformation tensor by,
\[
{\bf r}_{\alpha i} ({\bf R}_i, t)
:= {\bfg \phi}_{\alpha} (t) \cdot {\bf S}_{\alpha i}(t),
\]
where ${\bf S}_{\alpha i} (t)$ is a statistical field variable
that is a function of time. This is in contrast to
the deterministic continuum mechanics, in which ${\bf R}_{\alpha i}$
is a fixed vector in the referential configuration.
A simple way to distinguish the deterministic referential coordinate
${\bf R}_i$ and statistical coordinate ${\bf S}_i$ is
to set
\[
{\bf S}_i (t) = {\bf R}_i + {\bf T}_i (t),~~{\rm and}
~~<{\bf T}_i (t)> =0~,~~{\rm or}~~
<{\bf S}_i (t)> ={\bf R}_i,
\]
where the operator $< \cdot >$ is temporal average operator,
which will be discussed in Section 3 in details.

The Second Multiscale Decomposition is the Wagner and Liu's
additive decomposition,
in which the following additive decomposition
of the displacement field is introduced,
\begin{equation}
{\bf u}_i (t) =
\bar{\bf u}_i ({\bf R}_i,t) + {\bf u}^{\prime}_i (t)
\label{eq:WL-decomposition}
\end{equation}
where ${\bf u}^{\prime}_i$ is the fine scale displacement of
the i-th atom, and
$\bar{\bf u}_i$ is the value of the coarse scale displacement
of the atom $i$ that
may be determined by continuous displacement field, e.g.
the finite element interpolation field,
\[
\bar{\bf u} ({\bf R}_{i}) = \sum_I N_I ({\bf R})
\Bigm|_{{\bf R} = {\bf R}_{i}}
{\bf d}_I
\]
where $N_I ({\bf R})$ is the finite element interpolation
function.
Following this construction, intuitively
one may define and the fine scale relative displacement field
in a micromorphic displacement field as
\begin{equation}
{\bf r}^{\prime}_{\alpha i}= {\bfg \phi}_{\alpha} (t) {\bf S}_i -
{\bf F} ({\bf R}_{\alpha}, t) \cdot {\bf R}_{\alpha i}
\label{eq:fine-scale-2}
\end{equation}
where ${\bf F} ({\bf R}_{\alpha}, t) $ is the coarse scale deformation
gradient, which can be calculated as,
\[
{\bf F} ({\bf R}_{\alpha}, t) = {\partial {\bf r} \over
\partial {\bf R}}\Bigm|_{{\bf R} = {\bf R}_{\alpha}} = \sum_I
\Bigl(
{\bf d}_I \otimes {\partial N_I \over \partial {\bf R}}
\Bigr) \Bigm|_{{\bf R} = {\bf R}_{\alpha}} + {\bf I}^{(2)}
\]
This definition (\ref{eq:fine-scale-2}) is consistent with
(\ref{eq:WL-decomposition}), e.g.
\[
{\bf r}_{\alpha i} = \bar{\bf r}_{\alpha i}
+ {\bf r}^{\prime}_{\alpha i},
\]
if only the Cauchy-Born rule can be applied
for the coarse scale displacement, i.e.
\[
\bar{\bf r}_{\alpha i} \approx {\bf F} ({\bf R}_{\alpha}, t)
\cdot {\bf R}_{\alpha i},~~{\rm and}~~{\bf F} ({\bf R}_{\alpha},0) =
{\bf I}^{(2)}~.
\]
In the rest of the paper,
we assume that the Cauchy-Born rule approximation stands
at coarse scale, unless it is stated otherwise.

\section{Statistical closures of PR-MD}
In the micromorphic multiscale decomposition,
the fine scale displacement is determined
by the following multiplicative decomposition,
\begin{equation}
{\bf r}^{\prime}_{\alpha i} = {\bf r}_{\alpha i} -
\bar{\bf r}_{\alpha i},~~
{\bf r}_{\alpha i}(t) := {\bfg \phi}_{\alpha} (t) \cdot {\bf S}_i (t),
\end{equation}
where ${\bfg \phi}_{\alpha}$ is the micro-deformation tensor
of the $\alpha$-th cell,
and the vector ${\bf S}_i$ belongs to a time-dependent configuration
that is overlap with $\Omega_{\alpha}(0)$, and ${\bf S}_{i}(0) = {\bf R}_{\alpha i}$.
Since,
\begin{equation}
\dot{\bf r}_{\alpha i} = \dot{\bfg \phi}_{\alpha} {\bf S}_i
+ {\bfg \phi}_{\alpha} \dot{\bf S}_i
\end{equation}
It is trivial to show that both
\begin{equation}
\sum_i m_i {\bf S}_i = 0,~~{\rm and}~~\sum_{i} m_i \dot{\bf S}_i = 0~.
\end{equation}

For simplicity we only
consider the following first-principle Lagrangian for the $\alpha$-th
cell with pair potential and usual external potential
\begin{equation}
\mathcal{L}_{\alpha} = {1\over 2} \sum_i m_i \dot{\bf r}_i \cdot \dot{\bf r}_i
- {1 \over 2} \sum_{i} \sum_{j\not =i} V (r_{ij}) - \sum_{i}
{\bf f}_i \cdot {\bf r}_i
\label{eq:FF1}
\end{equation}
where $\phi (r_{ij})$ is the pair potential, $r_{ij} = |{\bf r}_{ij}|$
and ${\bf r}_{ij} = {\bf r}_j - {\bf r}_i$; ${\bf f}_i$ is
the external force acting on the atom $i$.

By considering the center of mass decomposition,
the total kinetic energy becomes,
\begin{eqnarray}
\mathcal{K} &=& {1 \over 2} \sum_{i} m_i
(\dot{\bf r}_{\alpha} + \dot{\bfg \phi}_{\alpha} \cdot {\bf S}_i
+ {\bfg \phi}_{\alpha} \cdot \dot{\bf S}_i )
\nonumber
\\
&{}&\cdot
(\dot{\bf r}_{\alpha} + \dot{\bfg \phi}_{\alpha} \cdot {\bf S}_i
+ {\bfg \phi}_{\alpha} \cdot \dot{\bf S}_i )
\nonumber
\\
&=& \underbrace{
{M \over 2} \dot{\bf r}_{\alpha} \cdot \dot{\bf r}_{\alpha}
}_\text{$\mathcal{K}_1$}
+
\underbrace{
{1 \over 2} \sum_{i} m_i {\bf S}_i \cdot
(\dot{\bfg \phi}_{\alpha}^{T}\dot{\bfg \phi}_{\alpha}) \cdot {\bf S}_i
}_\text{$\mathcal{K}_2$}
\nonumber
\\
&+&
\underbrace{
{1 \over 2} \sum_i m_i \dot{\bf S}_i \cdot {\bf C}
\cdot \dot{\bf S}_i
}_\text{$\mathcal{K}_3$}
\nonumber
\\
&+&
\underbrace{
{1 \over 2} \sum_i m_i \left (
{\bf S}_i (\dot{\bfg \phi}_{\alpha}^{T} {\bfg \phi}_{\alpha})
\cdot \dot{\bf S}_i
+ \dot{\bf S}_i ({\bfg \phi}^{T}_{\alpha}
\dot{\bfg \phi}_{\alpha}) {\bf S}_i \right )
}_\text{$\mathcal{K}_4$}
\label{eq:eqK}
\end{eqnarray}

The second term of (\ref{eq:eqK}) may be written as
\begin{equation}
\mathcal{K}_2=
{1 \over 2} (\dot{\bfg \phi}^{T}_{\alpha}
\dot{\bfg \phi}_{\alpha}) \sum_i m_i {\bf S}_i \otimes
{\bf S}_i~.
\end{equation}
To link the miscroscale quantities to macroscale field
variables,
we introduce the following statistical assumption,
\begin{equation}
\sum_i m_i {\bf S}_i \otimes {\bf S}_i =
I_{IJ} {\bf E}_I \otimes {\bf E}_J = const.
\end{equation}
which we coined as the first Parrinello-Rahman closure condition,
and it is obviously valid under equilibrium states.

Let
\begin{equation}
\dot{\phi}^{\alpha}_{iI} \dot{\phi}^{\alpha}_{iJ} = \omega_{I} \omega_J
\end{equation}
The second term of the kinetic energy may be written as
\begin{equation}
{1 \over 2} {\bfg \omega} \cdot {\bf J} \cdot {\bfg \omega}
= {1 \over 2} J_{IJ} \omega_{I} \omega_J
\end{equation}
Note that ${\bf J}$ is not the standard moment inertia tensor.
The standard moment inertia tensor is defined as
\[
{\bf J}_s = \int_{V} \rho (\|{\bf R}\|^2 {\bf I}^{(2)}
- {\bf R} \otimes {\bf R} ) d V
\]
where ${\bf I}^{(2)}$ is the second order unit tensor.

If $\{ {\bf E}_I \}$ are principal axes, we have
\[
{\bf J} = J_{11} {\bf E}_1 \otimes {\bf E}_1
+ J_{22} {\bf E}_2 \otimes {\bf E}_2
+J_{33} {\bf E}_3 \otimes {\bf E}_3
\]
If $J_{11}=J_{22}=J_{33}=W$, we have the following result
from Parrinello and Rahman \cite{PR81},
\begin{equation}
\mathcal{K}_2 = {1 \over 2} W tr (\dot{\bfg \phi}^{T}_{\alpha}
\dot{\bfg \phi}_{\alpha})~.
\end{equation}
In practice, one may make the following approximation,
\begin{eqnarray}
{\bf J} &=& \sum_i m_i {\bf S}_i \otimes {\bf S}_i
\approx \sum_i m_i {\bf S}_i (0) \otimes {\bf S}_i (0)
\nonumber
\\
&=& \sum_i m_i {\bf R}_{\alpha i} \otimes {\bf R}_{\alpha i}
\end{eqnarray}

In \cite{PR81}, Parrinello and Rahman made another choice, and they let
\begin{eqnarray}
\mathcal{K}_4 &=& {1 \over 2} \sum_i m_i \left (
{\bf S}_i (\dot{\bfg \phi}_{\alpha}^{T} {\bfg \phi}_{\alpha})
\cdot \dot{\bf S}_i
+ \dot{\bf S}_i ({\bfg \phi}^{T}_{\alpha}
\dot{\bfg \phi}_{\alpha}) {\bf S}_i \right )  =0~,
\label{eq:eqK2}
\end{eqnarray}
which we call as the second Parrinello-Rahman closure.

In this paper, we shall not adopt
the second Parrinello-Rahman closure,
and instead we shall examine
the detailed contribution to the Euler-Lagrange
equations from the term $\mathcal{K}_5$.
First, we may write,
\[
\mathcal{K}_4 =
tr \left \{
\dot{\bfg \phi}_{\alpha} \cdot \Bigl(
\sum_i m_i {\bf S}_i \otimes \dot{\bf S}_i
\Bigr) \cdot {\bfg \phi}^{T}_{\alpha}
\right \}
\]

We can then calculate the following partial derivatives,
\begin{eqnarray}
{\partial \mathcal{K}_4 \over \partial {\bf S}_i}
&=& \dot{\bfg \phi}^{T}_{\alpha}
{\bfg \phi}_{\alpha} \cdot \Bigl(
\sum_i m_i \dot{\bf S}_i
\Bigr) = 0;
\label{eq:DD1}
\\
{\partial \mathcal{K}_4 \over \partial \dot{\bf S}_i}
&=& {\bfg \phi}^{T}_{\alpha}
\dot{\bfg \phi}_{\alpha} \cdot \Bigl(
\sum_i m_i {\bf S}_i
\Bigr) =0;
\label{eq:DD2}
\\
{\partial \mathcal{K}_4 \over \partial {\bfg \phi}_{\alpha}}
&=&
\dot{\bfg \phi}_{\alpha} \cdot
\Bigl( \sum_i m_i {\bf S}_i \otimes \dot{\bf S}_i
\Bigr)
\label{eq:DD3}
\\
{\partial \mathcal{K}_4 \over \partial \dot{\bfg \phi}_{\alpha}}
&=&
{\bfg \phi}_{\alpha} \cdot
\Bigl( \sum_i m_i \dot{\bf S}_i \otimes {\bf S}_i
\Bigr)
\label{eq:DD4}
\end{eqnarray}
If we choose the following statistical closures,
\begin{eqnarray}
\Bigl( \sum_i m_i {\bf S}_i \otimes \dot{\bf S}_i \Bigr) = 0,~&(a)&~
\nonumber
\\
\Bigl( \sum_i m_i \dot{\bf S}_i \otimes {\bf S}_i
\Bigr)=0~, &(b)&
\label{eq:EE2}
\end{eqnarray}
we shall have
$\displaystyle {\partial \mathcal{K}_4 \over \partial {\bfg \phi}_{\alpha}} = 0$
and
$\displaystyle {\partial \mathcal{K}_4 \over \partial \dot{\bfg \phi}_{\alpha}} = 0$.

In fact, the first Parrinello-Rahman closure implies that
\begin{equation}
\Bigl( \sum_i m_i {\bf S}_i \otimes \dot{\bf S}_i \Bigr) +
\Bigl( \sum_i m_i \dot{\bf S}_i \otimes {\bf S}_i
\Bigr)=0,
\label{eq:EE3}
\end{equation}
but obviously Eq. (\ref{eq:EE2})(a) and (b) are stronger
conditions than (\ref{eq:EE3}). Thus the closure conditions
(\ref{eq:EE2})(a) and (b) imply the first
Parrinello-Rahman closure.
In this paper, we call (\ref{eq:EE2})(a) and (b)
as the mesoscale closure conditions.

Since $\mathcal{K}_4$ may be re-written as the following form
\begin{equation}
\mathcal{K}_4 = \sum_i {m_i \over 2}
\left \{
(\dot{\bfg \phi}^{T}_{\alpha}{\bfg \phi}_{\alpha}):
({\bf S}_i \otimes \dot{\bf S}_i) +
({\bfg \phi}^{T}_{\alpha}\dot{\bfg \phi}_{\alpha}):
(\dot{\bf S}_i \otimes {\bf S}_i)
\right \},
\end{equation}
it is clear that the mesoscale closure can effectively enforce
both the first and the second Parrinello-Rahman
closures.
To justify the above statistical closures,
we study the correlation property of
the random variables $\{ {\bf S}_i \}$, which
have the property,
\[
\sum_i {\bf S}_i = 0~.
\]

We first define the following tensorial autocorrelation
function,
Consider the following tensorial correlation function,
\begin{eqnarray}
{\bf AC}_1(\tau) &=& {1 \over 2}
\Bigl( < {\bf S}(t) \otimes {\bf S} (t+\tau)>
+
< {\bf S}(t+\tau) \otimes {\bf S} (t)>
\Bigr)
\nonumber
\\
&:=&{1 \over 2N}
\Bigl( \sum_{i} m_i {\bf S}_i (t) \otimes {\bf S}_i (t+ \tau)
+  \sum_{i} m_i {\bf S}_i (t+\tau) \otimes {\bf S}_i (t)~.
\Bigr)
\end{eqnarray}
It is obvious that ${\bf AC}_1 (\tau)$ is an even function, and
\[
{d \over d \tau} {\bf AC}_1 \Bigm|_{\tau =0} = 0,~~\to~~
\sum_{i} m_i \dot {\bf S}_i (t) \otimes {\bf S}_i (t)
+  \sum_{i} m_i {\bf S}_i (t) \otimes \dot {\bf S}_i (t) = 0~,
\]
which is the statement of the first Parrinello-Rahman closure.

Next we consider the following tensorial autocorrelation function,
\begin{eqnarray}
{\bf AC}_2(\tau) &=& < {\bf S}(t) \otimes {\bf S} (t+\tau)>
= {1 \over 2}
\Bigl( < {\bf S}(t) \otimes {\bf S} (t+\tau)>
+
< {\bf S}(t-\tau) \otimes {\bf S} (t)>
\Bigr)
\nonumber
\\
&:=&{1 \over 2N}
\Bigl( \sum_{i} m_i {\bf S}_i (t) \otimes {\bf S}_i (t+ \tau)
+  \sum_{i} m_i {\bf S}_i (t-\tau) \otimes {\bf S}_i (t)
\Bigr)
\end{eqnarray}

If we assume that the autocorrelation is a constant or slowly changing
tensor in the neighborhood of $\tau = 0$, we can then have
\begin{equation}
{d \over d \tau} {\bf AC}_2 \Bigm|_{\tau =0} = 0~~\to~~
 \sum_{i} m_i \dot{\bf S}_i (t) \otimes {\bf S}_i (t)
= \sum_{i} m_i {\bf S}_i (t) \otimes \dot{\bf S}_i (t) = 0~,
\end{equation}
which is the mesoscale closure.

\section{Andersen-Parrinello-Rahman Molecular Dynamics}
As Ray and Rahman \cite{Ray84} referred it as $N {\sigma} H$ ensemble MD,
the anisotropic Lagragian of
the Andersen-Parrinello-Rahman molecular dynamics
for an isoenthalpic-isobaric ensemble is,
\begin{eqnarray}
\mathcal{L}_{\alpha}
&=&
{1 \over 2} \sum_{i} m_i {\bf S}_i \cdot
(\dot{\bfg \phi}_{\alpha}^{T}\dot{\bfg \phi}_{\alpha}) \cdot {\bf S}_i
\nonumber
\\
&+&
{1 \over 2} \sum_i m_i \dot{\bf S}_i \cdot {\bf C}
\cdot \dot{\bf S}_i
\nonumber
\\
&-& {1 \over 2} \sum_{i} \sum_{j\not =i} V (r_{ij}) - \sum_{i}
{\bf f}_i \cdot {\bf r}_i
 - \tilde{\bf P}: {\bfg \phi}_{\alpha} \Omega_{\alpha}
\end{eqnarray}
where $\tilde{\bf P}$ is the thermodynamic conjugate
stress measure of ${\bfg \phi}_{\alpha}$.

The basic equations of the PR molecular dynamics are
\begin{eqnarray}
&{}&{d \over d t} {\partial \mathcal{L}_{\alpha} \over
\partial \dot {\bfg \phi}_{\alpha}} -
{\partial \mathcal{L}_{\alpha} \over
\partial {\bfg \phi}_{\alpha}} = 0,
\\
&{}&{d \over d t} {\partial \mathcal{L}_{\alpha} \over
\partial \dot {\bf S}_{i}} -
{\partial \mathcal{L}_{\alpha} \over
\partial {\bf S}_{i}} = 0~.
\end{eqnarray}
Following the standard procedure, one may derive
the equations of motions for the multiscale molecular
dynamics in the $\alpha$-th cell,
\begin{eqnarray}
&{}&\ddot{\bf S}_i = - \sum_{j\not =i}
\Bigl( { V^{'} (r_{ij}) \over m_i r_{ij}} \Bigr)
( {\bf S}_i - {\bf S}_j)  - {\bf C}^{-1} \cdot \dot{\bf C} \cdot
\dot{\bf S}_i
 - {\bfg \phi}^{-1}_{\alpha} \cdot {\bf f}_i \otimes {\bf S}_i,~
{\rm and}
\label{eq:GG2}
\\
&{}& \ddot{\bfg \phi}_{\alpha} \cdot {\bf J}
= J {\bfg \sigma}_{\alpha} {\bfg \phi}^{-T}_{\alpha} \Omega_0 - \tilde{\bf P}
\Omega_{\alpha 0}
- \sum_i {\bf f}_i \otimes {\bf S}_i
\label{eq:GG3}
\end{eqnarray}
where ${\bfg \sigma}_{\alpha}$ is the Virial stress defined as
\begin{equation}
{\bfg \sigma}_{\alpha} =
{1 \over \Omega_{\alpha}} \sum_i
\Bigl(
- m_i {\bf v}^{\prime}_i \otimes {\bf v}^{\prime}_i
+ \sum_{j\not =i}
V^{'} (r_{ij})\Bigl(
{ {\bf S}_{ij} \otimes {\bf S}_{ij} \over
r_{ij} }
\Bigr)  \Bigr)
\end{equation}
where $\Omega_{\alpha} (t)$ is the volume of the $\alpha$-th MD cell,
and the fine scale velocity is defined as
\begin{equation}
{\bf v}^{\prime}_i = {\bfg \phi}_{\alpha} \cdot \dot{\bf S}_i
\end{equation}
One can verify that indeed
\[
\sum_i m_i {\bf v}^{\prime}_i = 0~.
\]
That is ${\bf v}^{\prime}_i$ is a peculiar velocity, but
${\bf v}^{\prime}_i \not = \dot{\bf r}_{\alpha i}$.

Based on nonlinear continuum mechanics \cite{Marsden-Hughes},
one may define the first Piola-Kirchhoff stress as,
\begin{equation}
{\bf P}_{\alpha} = J {\bfg \sigma}_{\alpha} {\bfg \phi}^{-1}_{\alpha}
\end{equation}
so the third equation becomes
\begin{equation}
\ddot{\bfg \phi}_{\alpha} \cdot {\bf J}
= \bigl( {\bf P}_{\alpha} - \tilde{\bf P} \bigr) \Omega_{\alpha 0}
- \sum_i {\bf f}_i \otimes {\bf S}_i
\end{equation}
In above derivations, Eq. (\ref{eq:GG2})-(\ref{eq:GG3})
are essentially the same as
those of Parrinello and Rahman's original formulation except
the external potential energy. To this end, we have
present a first-principle based justification
of APR molecular dynamics.

For APR-MD, in the original single cell of atom ensemble,
the position of the center of mass is
fixed, because the periodic boundary condition is
used. Hence
\[
{\partial \over \partial {\bf r}_{\alpha}} V(r_{ij}) = 0~.
\]
In this case, the centers of mass of different cells only undergo
rigid motion.

In coupling of multiple different cells without imposing
the periodic boundary condition,
the relative displacements of each center of mass
will be different, so that
\[
{\partial \over \partial {\bf r}_{\alpha}} V(r_{ij}) \not = 0~.
\]
In the next Sections,
we shall discuss the multiscale micromorphic molecular dynamics
that is applicable to arbitrary domain with arbitrary boundary
conditions.

\section{Multiscale micromorphic molecular dynamics}
In this Section, we shall extend the periodic boundary condition
based APR-MD to an arbitrary finite size multiscale
micromorphic molecular dynamics.

To extend APR molecular dynamics to mesoscale scale and continuum
scale with arbitrary boundary conditions, we propose
the following three scale kinematic decomposition,
\begin{eqnarray}
{\bf r}_i &=& {\bf r}_{\alpha} + {\bfg \phi}_{\alpha} \cdot {\bf S}_i
\label{eq:MMMD1}
\\
{\bfg \phi}_{\alpha} &=& {\bfg \chi}_{\alpha} \cdot {\bf F}_{\alpha}
\label{eq:MMMD2}
\end{eqnarray}
where ${\bf r}_i$ is the position of the i-th atom of the system in the deformed
configuration;  ${\bf S}_i$
the random vibration position of the i-th atom in the referential configuration;
${\bf r}_{\alpha}$ is the center of mass of $\alpha$-th unit cell;
${\bfg \chi}_{\alpha}$ is the micro deformation of the $\alpha$-th unit cell,
and ${\bf F}_{\alpha}$ is the coarse scale deformation gradient, which is
determined by the relative position of the centers of mass of different cells.
The three independent kinematic variables at three different scales are:
$\{  {\bf S}_i, {\bfg \chi}_{\alpha}, {\rm and}~{\bf r}_{\alpha} \}$.
The novelty of the proposed multiscale decomposition is the multiplicative
multiscale decomposition Eq. (\ref{eq:MMMD2}).
We note that even though the coarse scale deformation gradient only depends
on the relative position of the centers of mass of different cells, i.e.
${\bf F}_{\alpha} = {\bf F}_{\alpha} (\{ {\bf r}_{\beta} \})$,
it may take different values for different atoms in a same cell, i.e.
${\bf F}_{\alpha} ({\bf R}_i, \{{\bf r}_{\beta}\} ) \not =
{\bf F}_{\alpha} ({\bf R}_j, \{{\bf r}_{\beta}\} ),
~ {\bf R}_i, {\bf R}_j \in \Omega_{\alpha 0}$.
More precisely, Eqs. (\ref{eq:MMMD1}) and (\ref{eq:MMMD2}) may be written as
\begin{eqnarray}
{\bf r}_i &=& {\bf r}_{\alpha} + {\bfg \phi}_{\alpha} ({\bf R}_i) \cdot {\bf S}_i
\label{eq:MMMD3}
\\
{\bfg \phi}_{\alpha} ({\bf R}_i) &=& {\bfg \chi}_{\alpha}
\cdot {\bf F}_{\alpha} ({\bf R}_i, \{{\bf r}_{\beta}\})
\label{eq:MMMD4}
\end{eqnarray}
with the understanding that ${\bf R}_i$ is just an interpolation variable,
and ${\bf F}_{\alpha} = {\bf F}_{\alpha} ({\bf R}_{i})$
does not depend on dynamics variable ${\bf S}_i$.
In the subsequent Sections,
we shall discuss the difference between the case
that ${\bf F}_{\alpha}$ depends on ${\bf R}_i$ in a fixed cell and
the case that
${\bf F}_{\alpha}$ is a constant tensor in the entire cell.
If we denote the lattice spacing as $\ell_a$, the unit cell size as $\ell_c$, and
the nonlocal support of a center of mass particle as $\ell_r$, we have
\[
\ell_a < \ell_c < \ell_r~~.
\]

The time derivatives of the independent kinematic variables are,
\begin{eqnarray}
\dot{\bf r}_{\alpha} &=& \dot{\bf r}_{\alpha} +
\dot{\bfg \phi}_{\alpha} \cdot {\bf S}_i + {\bfg \phi}_{\alpha} \cdot \dot{\bf S}_i
\nonumber
\\
\dot{\bfg \phi}_{\alpha} &=& \dot{\bfg \chi}_{\alpha} \cdot
{\bf F}_{\alpha} + {\bfg \chi}_{\alpha} \cdot \dot{\bf F}_{\alpha}
\nonumber
\end{eqnarray}
If we denote the time scale of $\dot{\bf S}_i$ as $t_s$, the time scale of $\dot{\bfg \chi}_{\alpha}$
as $t_c$, and time scale $\dot{\bf r}_{\alpha}$ ad $t_r$, we again have
\[
t_s < t_c < t_r~.
\]

We start with a first principle Lagrange (\ref{eq:FF1})
of a multiscale ensemble (multiple cells) system
in terms of following multiscale decomposition,
\begin{eqnarray}
\mathcal{L}_m
&=& {1 \over 2} \sum_{\beta} M_{\beta} \dot{\bf r}_{\beta}
\cdot \dot{\bf r}_{\beta}
+ {1 \over 2} \sum_{\beta} {\bf J}_{\beta} :
(\dot{\bfg \phi}_{\beta}^{T}\dot{\bfg \phi}_{\beta})
\nonumber
\\
&+& {1 \over 2} \sum_{\beta}
\sum_i m_i \dot{\bf S}_i \cdot {\bf C}_{\beta}
\cdot \dot{\bf S}_i
- {1 \over 2} \sum_{\beta } \sum_{\gamma}
\sum_{i \in \beta, j \in \gamma} V (r_{ij})
\nonumber
\\
&&- \sum_{\beta} \sum_{i \in \beta} {\bf f}_i
\cdot {\bf r}_{i}
\end{eqnarray}
where $\beta, \gamma$ are cell indices, and
the abbreviation $i \in \beta $ means that the i-th atom
belongs to the $\beta$-th cell.
${\bf C}_{\beta} :={\bfg \phi}^{T}_{\beta} {\bfg \phi}_{\beta}$
 is the micro right Cauchy-Green tensor for total deformation. We denote
 that
$\displaystyle M_{\beta} = \sum_{i \in \beta} m_i$ and
$\mathcal{B}_{\beta} = \displaystyle \sum_{i \in \beta} {\bf f}_i $.
In the rest of this paper, we always assume that
the Roman index $i$ is used to make the atoms inside
the $\beta$-th cell, whereas the Roman index $j$ is designated
to denote the atoms inside the $\gamma$-th cell.

For simplicity, we choose three independent field variables,
${\bf r}_{\alpha}, {\bfg \phi}_{\alpha}$,
and ${\bf S}_i$ for three scales, i.e.
$\mathcal{L}_m = \mathcal{L}_m ({\bf r}_{\alpha}, {\bfg \phi}_{\alpha},{\bf S}_i)$
We postulate the following principle of multiscale stationary action,
\[
\delta \mathcal{S}
[{\bf r}_{\alpha}, {\bfg \phi}_{\alpha}, \{ {\bf S}_i \}]
 = \int_{t_1}^{t_2} \delta \mathcal{L}_m
(\dot{\bf r}_{\alpha}, \dot{\bfg \phi}_{\alpha}, \{ \dot {\bf S}_i \} ,
{\bf r}_{\alpha}, {\bfg \phi}_{\alpha}, \{ {\bf S}_i \}, t )d t = 0~.
\]
The Lagrangian equations of
the multiscale micromorphic molecular dynamics can be derived as follows,
\begin{eqnarray}
&{}&{d \over d t} {\partial \mathcal{L}_{m} \over
\partial \dot {\bf r}_{\alpha}} -
{\partial \mathcal{L}_{m} \over
\partial {\bf r}_{\alpha}} = 0,
\\
&{}&{d \over d t} {\partial \mathcal{L}_{m} \over
\partial \dot {\bfg \phi}_{\alpha}} -
{\partial \mathcal{L}_{\alpha} \over
\partial {\bfg \phi}_{\alpha}} = 0,
\\
&{}&{d \over d t} {\partial \mathcal{L}_{m} \over
\partial \dot {\bf S}_{i}} -
{\partial \mathcal{L}_{\alpha} \over
\partial {\bf S}_{i}} = 0~.
\end{eqnarray}
Note that one may choose ${\bfg \phi}_{\alpha}$ instead of ${\bfg \chi}_{\alpha}$
as the independent field variable
for mesoscale dynamic equations. Since
${\bfg \phi}_{\alpha} = {\bfg \chi}_{\alpha} \cdot {\bf F}_{\alpha}$,
the two choices will be equivalent.

\medskip
\subsection{Coarse scale dynamic equations}

We start by deriving some useful relations that are needed in
the subsequent derivations.
Since we know that
\begin{eqnarray}
{\bf F}_{\beta} = {\bf F}_{\beta}(\{{\bf r}_{\alpha}\}) ,  \ \ {\bf \dot F}_{\beta} = {\bf \dot F}_{\beta}(\{{\bf r}_{\alpha}\}, \{{\bf \dot r}_{\alpha}\})
\end{eqnarray}
then we have
\begin{eqnarray}
{\bf \dot F}_{\beta} = \sum_{\alpha}\frac{\partial {\bf F}_{\beta}}{\partial {\bf r}_{\alpha}}{\bf \dot r}_{\alpha},
\end{eqnarray}
which leads to the relation,
\begin{eqnarray}
\frac{\partial {\bf \dot F}_{\beta}}{\partial {\bf \dot r}_{\alpha}}
= \frac{\partial {\bf F}_{\beta}}{\partial {\bf r}_{\alpha}},
\label{fdd0}
\end{eqnarray}
and
\begin{eqnarray}
{\bf \ddot F}_{\beta} = \sum_{\alpha}
\left( \frac{d}{dt} \left(\frac{\partial {\bf F}_{\beta}}{\partial {\bf r}_{\alpha}}\right){\bf \dot r}_{\alpha}
+  \frac{\partial {\bf F}_{\beta}}{\partial {\bf r}_{\alpha}}{\bf \ddot r}_{\alpha}
 \right) ~.
 \label{fdd1}
\end{eqnarray}
On the other hand, we may derive,
\begin{eqnarray}
{\bf \ddot F}_{\beta} = \sum_{\alpha}\left(   \frac{\partial {\bf \dot F}_{\beta}}{\partial {\bf r}_{\alpha}}{\bf \dot r}_{\alpha}
+  \frac{\partial {\bf \dot F}_{\beta}}{\partial {\bf \dot r}_{\alpha}}{\bf \ddot r}_{\alpha}  \right)  \label{fdd2}
\end{eqnarray}
Comparing equations \eqref{fdd1} and \eqref{fdd2} and utilizing
 \eqref{fdd0}, we obtain
\begin{eqnarray}
 \frac{d}{dt} \left(\frac{\partial {\bf F}_{\beta}}{\partial {\bf r}_{\alpha}}\right)
 =  \frac{\partial {\bf \dot F}_{\beta}}{\partial {\bf r}_{\alpha}}  ~.
 \label{fdd3}
\end{eqnarray}
Next we are first taking time derivative on ${\bfg \phi}_{\beta}$,
\begin{eqnarray}
\dot{\bfg \phi}_{\beta} = \dot{\bfg \chi}_{\beta} {\bf F}_{\beta}
 + {\bfg \chi}_{\beta} \dot{\bf F}_{\beta},
\end{eqnarray}
and then we can find that
\begin{eqnarray}
\frac{  \partial \dot{\bfg  \phi}_{\beta}  }{  \partial \dot{\bf r}_{\alpha}  }= {\bfg \chi}_{\beta} \frac{\partial \dot{\bf F}_{\beta}}{\partial \dot{\bf r}_{\alpha}}
={\bfg \chi}_{\beta} \frac{\partial {\bf F}_{\beta}}{\partial {\bf r}_{\alpha}}=\frac{  \partial {\bfg  \phi}_{\beta}  }{  \partial {\bf r}_{\alpha}  }    ~.
  \label{pd1}
\end{eqnarray}
By virtue of Eqs. \eqref{fdd3} $\sim$ \eqref{pd1}, we have
\begin{eqnarray}
\frac{\partial \dot{\bfg \phi}_{\beta}}{\partial {\bf r}_{\alpha}} &=&
\dot{\bfg \chi}_{\beta} \frac{\partial {\bf F}_{\beta}}{\partial {\bf r}_{\alpha}} + {\bfg \chi}_{\beta} \frac{\partial \dot{\bf F}_{\beta}}{\partial {\bf r}_{\alpha}} \nonumber \\
&=&\dot{\bfg \chi}_{\beta} \frac{\partial {\bf F}_{\beta}}{\partial {\bf r}_{\alpha}}
+ {\bfg \chi}_{\beta}  \frac{d}{dt} \left(\frac{\partial {\bf F}_{\beta}}{\partial {\bf r}_{\alpha}}\right) \nonumber \\
&=& \frac{d}{dt} \left(     \frac{  \partial {\bfg  \phi}_{\beta}  }{  \partial {\bf r}_{\alpha}  }      \right)
= \frac{d}{dt} \left(     \frac{  \partial \dot{\bfg  \phi}_{\beta}  }{  \partial \dot{\bf r}_{\alpha}  }      \right)~.
\end{eqnarray}
This relation is needed in the subsequent derivation.

Reconsidering the Lagrangian equation at the coarse scale and utilizing the above relation,
we have
\begin{eqnarray}
{d \over d t} \Bigl(
{ \partial \mathcal{L}_m \over \partial \dot{\bf r}_{\alpha}}
\Bigr)
&=& {d \over d t}
\Bigl(
{\partial \mathcal{L}_m \over \partial \dot{\bf r}_{\alpha}}
+ \sum_{\beta} {\partial \mathcal{L}_m \over \partial \dot{\bfg \phi}_{\beta}}
\cdot {\partial \dot{\bfg \phi}_{\beta} \over \partial \dot{\bf r}_{\alpha}}
\Bigr)
\nonumber
\\
\nonumber
&=& M_{\alpha} \ddot{\bf r}_{\alpha}
+
\sum_{\beta} {d \over d t} \Bigl(
{\partial \mathcal{L}_m \over \partial \dot{\bfg \phi}_{\beta}}
\Bigr) \cdot {\partial \dot{\bfg \phi}_{\beta} \over \partial \dot{\bf r}_{\alpha}}
+
\sum_{\beta}
{\partial \mathcal{L}_m \over \partial \dot{\bfg \phi}_{\beta}}
 \cdot {d \over d t} \Bigl( {\partial \dot{\bfg \phi}_{\beta} \over \partial \dot{\bf r}_{\alpha}}\Bigr)
\nonumber
\\
&=&
M_{\alpha} \ddot{\bf r}_{\alpha}
+ \sum_{\beta} {\partial \mathcal{L}_m \over \partial {\bfg \phi}_{\beta}}
\cdot { \partial {\bfg \phi}_{\beta} \over \partial {\bf r}_{\alpha}}
+ \sum_{\beta} {\partial \mathcal{L}_m \over \partial \dot{\bfg \phi}_{\beta}}
\cdot {\partial \dot{\bfg \phi}_{\beta} \over
\partial {\bf r}_{\alpha}}
\label{eq:xxx1}
\end{eqnarray}
On the other hand,
\begin{eqnarray}
{\partial \mathcal{L}_m \over \partial {\bf r}_{\alpha}}
&=& {\partial \mathcal{L}_m \over \partial {\bf r}_{\alpha}}
+ \sum_{\beta} {\partial \mathcal{L}_m \over \partial {\bfg \phi}_{\beta}}
\cdot {\partial {\bfg \phi}_{\beta} \over \partial {\bf r}_{\alpha}}
+ \sum_{\beta}
{\partial \mathcal{L}_m \over \partial \dot{\bfg \phi}_{\beta}}
{\partial \dot{\bfg \phi}_{\beta} \over \partial {\bf r}_{\alpha}}
\nonumber
\\
&=& - \sum_{\beta \not =\alpha} \sum_{i \in \alpha, j \in \beta} V^{\prime} (r_{ij})
{{\bf r}_{ij} \over |{\bf r}_{ij}|} - \sum_{i \in \alpha} {\bf f}_i +
\sum_{\beta} {\partial \mathcal{L}_m \over \partial {\bfg \phi}_{\beta}}
\cdot {\partial {\bfg \phi}_{\beta} \over \partial {\bf r}_{\alpha}}
+ \sum_{\beta}
{\partial \mathcal{L}_m \over \partial \dot{\bfg \phi}_{\beta}}
\cdot
 {\partial \dot{\bfg \phi }_{\beta} \over \partial {\bf r}_{\alpha}}~.
 \label{eq:xxx2}
\end{eqnarray}
Combining Eqs. (\ref{eq:xxx1}) and (\ref{eq:xxx2}) and utilizing the coarse scale
Lagrangian equation,
\[
{d \over d t} \Bigl(
{\partial \mathcal{L}_m \over \partial \dot{\bf r}_{\alpha}}
\Bigr)
- {\partial \mathcal{L}_m \over \partial {\bf r}_{\alpha}} = 0~,
\nonumber
\]
we finally have
\begin{equation}
M_{\alpha} \ddot{\bf r}_{\alpha}
+ \sum_{\beta \not = \alpha}
\sum_{
 {i \in \alpha},
 {j \in \beta}
 }
 V^{\prime} (r_{ij})
{{\bf r}_{ij} \over |{\bf r}_{ij}|} + \mathcal{B}_{\alpha} = {\bf 0}~.
\end{equation}
It is clear that the first term is the coarse scale inertia term,
and the second term is the cell-cell interaction force, and the third term
is the external force acting on the center of mass of the $\alpha$-th cell.

\subsection{Mesoscale dynamic equations}
\noindent
Second, we exam the mesoscale Lagrangian equation,
\[
{d \over dt} {\partial \mathcal{L}_m \over \partial \dot{\bfg \phi}_{\alpha}}
- {\partial \mathcal{L}_m \over \partial {\bfg \phi}_{\alpha}} = 0~.
\]
For a systematic derivation, we denote
\begin{eqnarray}
&&{\bf r}_i = {\bf r}_{\beta} + {\bfg \phi}_{\alpha} \cdot {\bf S}_i,~~~{\rm and}~~
{\bf r}_j = {\bf r}_{\gamma} + {\bfg \phi}_{\gamma} \cdot {\bf S}_j~~\to
\nonumber
\\
&&{\bf r}_{ij} = {\bf r}_j - {\bf r}_i
=
{\bf r}_{\beta \gamma} + {\bfg \phi}_{\gamma} \cdot {\bf S}_j
- {\bfg \phi}_{\beta} \cdot {\bf S}_i
\end{eqnarray}
where
\[
{\bfg \phi}_{\beta} = {\bfg \chi}_{\beta} \cdot {\bf F}_{\beta}
~~{\rm and}~~
{\bfg \phi}_{\gamma} = {\bfg \chi}_{\gamma} \cdot
{\bf F}_{\gamma}
\]

To facilitate the subsequent derivation,
we first consider the derivative terms with respect to
the chosen mesoscale variable, i.e.
$\dot{\bfg \phi}_{\alpha}$ and ${\bfg \phi}_{\alpha}$ :\\
{\bf 1.}
\[
~~~~~~~~~~~~~~~~~~~~~~~
{\partial \mathcal{L}_m
\over \partial \dot{\bfg \phi}_{\alpha}} =
\dot{\bfg \phi}_{\alpha} \cdot {\bf J}_{\alpha}~~~~~~~~~~~~~~~~~~~~~~~~~~~~~~~~~
\]
where ${\bf J}_{\alpha} = \sum_i m_i {\bf S}_i \otimes {\bf S}_i \approx
\sum_i m_i {\bf R}_i  \otimes {\bf R}_i $.
Hence
\[
{d \over d t} \Bigl(
{\partial \mathcal{L}_m \over \partial \dot{\bfg \phi}_{\alpha}}
\Bigr)
= {d \over d t} ( \dot{\bfg \phi}_{\alpha} \cdot {\bf J}_{\alpha} )
= \ddot{\bfg \phi}_{\alpha} \cdot {\bf J}_{\alpha}
\]
{\bf 2.}
\[
{\partial {\bf C}_{\alpha} \over \partial {\bfg \phi}_{\alpha}}
= 2 {\bfg \phi}_{\alpha}~~~~~~~~~~~~~~~~~~~~~~~~~~~~~~~~~~~~~~~~
\]
{\bf 3.}
\begin{eqnarray}
&& (a)~\beta = \gamma:~~
{\partial r_{ij} \over \partial {\bfg \phi}_{\alpha}} =
{{\bf r}_{ij} \over r_{ij}} \cdot
{\partial {\bf r}_{ij} \over \partial {\bfg \phi}_{\alpha}}
 = \Bigl( {{\bf r}_{ij} \over r_{ij}} \otimes {\bf S}_{ij} \Bigr) \delta_{\alpha \beta}
\\
&& (b)~\beta \not = \gamma:~~
{\partial r_{ij} \over \partial {\bfg \phi}_{\alpha}} =
{{\bf r}_{ij} \over r_{ij}} \cdot
{\partial {\bf r}_{ij} \over \partial {\bfg \phi}_{\alpha}}
=  { {\bf r}_{ij} \over r_{ij}}
\otimes ( \delta_{\alpha \gamma} {\bf S}_j - \delta_{\alpha \beta} {\bf S}_i)~.
\end{eqnarray}

Hence,
\begin{eqnarray}
&& (a)~\beta = \gamma:~~
{\partial \mathcal{L}_m \over \partial
{\bfg \phi}_{\alpha}} = {1 \over 2} \sum_i
m_i \dot{\bf S}_i {\partial {\bf C}_{\alpha} \over \partial
{\bfg \phi}_{\alpha}}
\dot{\bf S}_i - {1 \over 2}
{\partial \over \partial {\bfg \phi}_{\alpha}}
 \sum_{\beta} \sum_{j \not = i, i,j \in \beta }V (r_{ij})
\nonumber
\\
&&~~~~~~~~=
{\bfg \phi}_{\alpha} \cdot \sum_i
m_i \dot{\bf S}_i \otimes \dot{\bf S}_i - {1 \over 2}
 \sum_{j \not = i, i,j \in \alpha }
{V^{\prime} (r_{ij}) \over r_{ij}} {\bfg \phi}_{\alpha} \cdot
 {\bf S}_{ij} \otimes {\bf S}_{ij}
\nonumber
\\
&& (b)~\beta \not = \gamma:~~
{\partial \mathcal{L}_m \over \partial
{\bfg \phi}_{\alpha}} =
{1 \over 2} \sum_i
m_i \dot{\bf S}_i {\partial {\bf C}_{\alpha} \over \partial
{\bfg \phi}_{\alpha}}
\dot{\bf S}_i - {1 \over 2}
{\partial \over \partial {\bfg \phi}_{\alpha}}
 \sum_{\beta} \sum_{\gamma} \sum_{i \in \beta, j \in \gamma } V (r_{ij})
\nonumber
\\
&&~~~~~~~~=
{\bfg \phi}_{\alpha} \cdot \sum_i
m_i \dot{\bf S}_i \otimes \dot{\bf S}_i - {1 \over 2}
\sum_{\beta} \sum_{\gamma} \sum_{ i \in \beta, j \in \gamma }
{V^{\prime} (r_{ij}) \over r_{ij}}
{\bf r}_{ij} \otimes
( \delta_{\alpha \gamma} {\bf S}_j - \delta_{\alpha \beta} {\bf S}_i)
\nonumber
\end{eqnarray}
The dynamic equations at mesoscale have the form,
\begin{eqnarray}
&&\ddot{\bfg \phi}_{\alpha} \cdot {\bf J}_{\alpha} - {\bfg \phi}_{\alpha}
\sum_i m_i \dot{\bf S}_i \otimes \dot{\bf S}_i
+
\sum_{ i, j \in \alpha, i \not = j }
{V^{\prime} (r_{ij}) \over r_{ij}}
{\bfg \phi}_{\alpha} \cdot {\bf S}_i \otimes {\bf S}_i
\nonumber
\\
&&+
\sum_{\beta \not = \gamma} \sum_{ i \in \beta, j \in \gamma }
{V^{\prime} (r_{ij}) \over r_{ij}}
{\bf r}_{ij} \otimes
\Bigl( \delta_{\alpha \gamma} {\bf S}_j - \delta_{\alpha \beta} {\bf S}_i \Bigr)
+ \sum_{i \in \alpha} {\bf f}_i \otimes {\bf S}_i = 0~.
\end{eqnarray}
where ${\bf r}_{ij} = {\bf r}_{\beta \gamma} + {\bfg \phi}_{\gamma} \cdot {\bf S}_j
- {\bfg \phi}_{\beta} \cdot {\bf S}_i$.

Define the mesoacle $2^{nd}$ Piola-Kirchhoff stress tensor,
\begin{eqnarray}
{\bf \mathcal{S}}_{\alpha}^{int} &:=& {1 \over \Omega_{\alpha 0}}
\sum_{i \in \alpha}
\Bigl(
- m_i \dot{\bf S}_i \otimes \dot{\bf S}_i
+
\sum_{j \in \alpha,  j \not = i }
{V^{\prime} (r_{ij}) \over r_{ij}}
{\bf S}_i \otimes {\bf S}_i \Bigr)
\label{eq:mesoEq1}
\\
{\bf \mathcal{S}}^{ext}_{\alpha} &=& {1 \over \Omega_{\alpha 0}}
\sum_{\beta \not = \alpha} \sum_{i \in \alpha, j \in \beta}
{V^{\prime} (r_{ij}) \over r_{ij}} {\bf r}_{ij} \otimes {\bf S}_i
\label{eq:mesoEq2}
\end{eqnarray}
where ${\bf r}_{ij} = {\bf r}_{\alpha \beta} + {\bfg \phi}_{\beta} \cdot {\bf S}_j
- {\bfg \phi}_{\alpha} \cdot {\bf S}_i$.

The mesoscale dynamics equations can be recast into
\[
\ddot{\bfg \phi}_{\alpha} \cdot {\bf J}_{\alpha} +
{\bfg \phi}_{\alpha} \cdot
\Bigl( {\bf \mathcal{S}}_{\alpha}^{int} -
{\bf \mathcal{S}}_{\alpha}^{ext}
\Bigr) \Omega_{\alpha 0}  + {\bf M}_{\alpha} = 0~,
\]
where $ \mathcal{\bf M}_{\alpha} =
\displaystyle \sum_{i \in \alpha} {\bf f}_i \otimes {\bf S}_i$ is
the mesoscale external couple.
Note that Eqs. (\ref{eq:mesoEq1}) and (\ref{eq:mesoEq2}) are insightful,
because it resolves one of outstanding debates on the definition of
the Virial Stress.
Eq. (\ref{eq:mesoEq1}) is basically the mathematical definition of
the Virial stress e.g. \cite{Irving50,Tsai79}.
However, Zhu \cite{Zhou03} argued that the kinetic energy
part should be dropped out in the stress calculation,
even though many disagreed, e.g. \cite{Murdoch07,Sun08}.
We now see from Eqs. (\ref{eq:mesoEq1}) and (\ref{eq:mesoEq2}) that
if the stress is internally generated, the definition of the virial
stress is the original definition of the virial stress; but if
the stress is an external stress, then the kinetic energy part should
drop out from its expression. This is because that the current formulation
of the multiscale micromorphic molecular dynamics
is formulated under adiabatic condition, which
does not consider the heat exchange among the cells.

\subsection{Microscale dynamic equations}
For simplicity, we re-index the multiscale Lagrangian as
\begin{eqnarray}
\mathcal{L}_m
&=& \sum_{\alpha} {M_{\alpha} \over 2} \dot{\bf r}_{\alpha}
\cdot \dot{\bf r}_{\alpha}
+ {1 \over 2} \sum_{\alpha} {\bf J}_{\alpha} :
(\dot{\bfg \phi}_{\alpha}^{T}\dot{\bfg \phi}_{\alpha})
\nonumber
\\
&+& {1 \over 2} \sum_{\alpha}
\sum_i m_i \dot{\bf S}_i \cdot {\bf C}_{\alpha}
\cdot \dot{\bf S}_i
- {1 \over 2} \sum_{\alpha } \sum_{\beta} \sum_{i\not=j} V (r_{ij})
\nonumber
\\
&-& \sum_{\alpha} \sum_{i} {\bf f}_i
\cdot {\bfg \phi}_{\alpha}\cdot  {\bf S}_{i}
- \sum_{\alpha} {\bf \mathcal{B}}_{\alpha} \cdot {\bf r}_{\alpha}
\end{eqnarray}
where the microscale variable ${\bf S}_i, i \in \alpha$
and ${\bf S}_j, j \in \beta$.

\begin{eqnarray}
&& (a)~\alpha = \beta,~~{\bf r}_{ij}
= {\bfg \phi}_{\alpha} \cdot {\bf S}_{ij},~~
{\partial r_{ij} \over \partial {\bf S}_{i}} =
- { {\bf r}_{ij} \over r_{ij}} \cdot {\bfg \phi}_{\alpha}
= - {  {\bf C}_{\alpha} \cdot {\bf S}_{ij} \over r_{ij}}
\\
&& (b)~\alpha \not = \beta:~~
{\bf r}_{ij} = {\bf r}_{\alpha \beta}
+ ({\bfg \phi}_{\beta} \cdot {\bf S}_j - {\bfg \phi}_{\alpha} \cdot {\bf S}_i),
~~
{\partial r_{ij} \over \partial {\bf S}_{i}} = -
{{\bf r}_{ij} \over r_{ij}} \cdot  {\bfg \phi}_{\alpha}.
\end{eqnarray}
Evaluating the fine scale Lagrangian equation for $i \in \alpha$,
\[
{d \over d t} {\partial \mathcal{L}_m \over \partial \emph{}\dot{\bf S}_i}
- {\partial \mathcal{L}_m \over \partial {\bf S}_i} = 0,~~i \in \alpha
\]
we have
\[
{d \over d t} {\partial \mathcal{L}_m \over \partial \dot{\bf S}_i}
= m_i \Bigl(
{\bf C}_{\alpha} \ddot{\bf S}_i + \dot{\bf C}_{\alpha}
\cdot \dot{\bf S}_i
\Bigr)
\]
and
\begin{eqnarray}
&& (a)~\alpha = \beta:~~
{\partial \mathcal{L}_m \over \partial {\bf S}_i}
= - {1 \over 2} \sum_{j \not = i}
\Bigl(
{ V^{\prime} (r_{ij}) \over r_{ij}} {\bf C}_{\alpha} \cdot
{\bf S}_{ij}
\Bigr)
\nonumber
\\
&& (b)~\alpha \not = \beta:~~
{\partial \mathcal{L}_m \over \partial {\bf S}_i}
= - {1 \over 2} \sum_{\alpha \not = \beta} \sum_{j \not = i}
\Bigl(
{ V^{\prime} (r_{ij}) \over r_{ij}} {\bfg \phi}^T_{\alpha} \cdot
{\bf r}_{ij}
\Bigr)
\end{eqnarray}
where ${\bf r}_{ij} = {\bf r}_{\alpha \beta} +
  {\bfg \phi}_{\beta} \cdot {\bf S}_j
  - {\bfg \phi}_{\alpha} \cdot {\bf S}_i $.

Finally, we can express the fine scale dynamics equations as,
\begin{eqnarray}
&& (a)~\alpha = \beta:~~
\ddot{\bf S}_i =
- {1 \over 2} \sum_{j\not = i}\Bigl(
{ V^{\prime} (r_{ij}) \over r_{ij}}
{\bf S}_{ij}
\Bigr) - {\bf C}^{-1}_{\alpha}
\dot{\bf C}_{\alpha} \cdot \dot{\bf S}_i
\\
&& (b)~\alpha \not = \beta:~~
\ddot{\bf S}_i =
- {1 \over 2} {\bfg \phi}^{-1}_{\alpha} \sum_{\alpha \not = \beta} \sum_{i \not = j}
\Bigl(
 {V^{\prime} (r_{ij}) \over r_{ij}}
  ( {\bf r}_{\alpha \beta} +
  {\bfg \phi}_{\beta} \cdot {\bf S}_j
  - {\bfg \phi}_{\alpha} \cdot {\bf S}_i)
\Bigr)
 - {\bf C}^{-1}_{\alpha}
\dot{\bf C}_{\alpha} \cdot \dot{\bf S}_i
\end{eqnarray}
Combining the two equations, we finally have
\begin{equation}
\ddot{\bf S}_i +
 {1 \over 2} {\bfg \phi}^{-1}_{\alpha} \sum_{\beta} \sum_{i \not = j}
\Bigl(
 {V^{\prime} (r_{ij}) \over r_{ij}}
  ( {\bf r}_{\alpha \beta} +
  {\bfg \phi}_{\beta} \cdot {\bf S}_j
  - {\bfg \phi}_{\alpha} \cdot {\bf S}_i)
\Bigr)
 + {\bf C}^{-1}_{\alpha}
\dot{\bf C}_{\alpha} \cdot \dot{\bf S}_i + {\bfg \phi}_{\alpha}^{-1}
\cdot {\bf f}_i = 0~.
\end{equation}
where $ i \in \alpha$.

\section{COARSE GRAINED MOLECULAR DYNAMICS}
By now, we have derived the exact and complete governing equations for
a three-scale micromorphic molecular dynamics,
which are based on the first principle Lagrangian.
This novel multiscale structure is an intrinsic property
of the original molecular dynamics. The only extrinsic parameter
is the size of the cell.

As one can find that the motions each scale are strongly
coupled to the others.
Thus we are able to couple them seamlessly.
One the other hand, different from most of the multiscale methods
that have been developed in recent years, whose main
purpose and advantages are the reduction of computation cost,
the proposed MMMD method is a different multiscale paradigm.
First, the MMMD is actually more complex and expensive than the original MD,
because we have to three sets of equations in three different
scales concurrently, and the time integration or the time scale
for three sets of dynamics equations are the same.
However, the expense of this complexity will allow us to use
MD as a nanomechanics tool to simulate finite size problems with
arbitrary boundary conditions.

Moreover, the multiscale micromorphic molecular dynamics discovered
in this work provides the theoretical foundation for us to derive or
to construct the coarse-grained molecular dynamics.
For instance, we can shut off molecular dynamics in
one or two scales to perform a single scale fast computation.
To illustrate this point,
we demonstrate in the following how to construct
a coarse-scale molecular dynamics.

We first propose to adopt
the Reproducing Kernel Particle Method \cite{Li-Liu99}
or the state-based Peridynamics \cite{Silling2} techniques
in construct the discrete deformation gradient.

To define the coarse scale representation,
we choose the coordinates of the center of mass of the each
cell as the coarse scale degrees of freedom, so that
the first principle Lagrange (\ref{eq:FF1}) is the
multiscale Lagrange without the need of further modification.
However, since the independent variable in
the coarse scale is the position of center of mass, ${\bf r}_{\alpha}$,
and we must link the coarse scale deformation gradient with
${\bf r}_{\alpha}$.
This can be done by employing an approach adopting
by the reproducing kernel particle method or
the state-based peridynamics \cite{Silling2},
in which the discrete deformation gradient is constructed as,
\begin{equation}
\mathbf{F}_{\alpha} = \Bigl( \sum_{\beta=1}^{N_h}
\omega (|{\bf R}_{\alpha \beta}|)
{\bf r}_{\alpha \beta} \otimes {\bf R}_{\alpha \beta} \Delta V_{\beta}
\Bigr) \cdot {\bf K}^{-1}_{\alpha}
\label{eq:PRD1}
\end{equation}
where ${\bf R}_{\alpha \beta} := {\bf R}_{\beta} - {\bf R}_{\alpha}$;
${\bf r}_{\alpha \beta} = {\bf r}_{\beta} - {\bf r}_{\alpha}$, and
\begin{equation}
{\bf K}_{\alpha} :=
\sum_{\beta=1}^{N_h}
\omega (|{\bf R}_{\alpha \beta}|)
{\bf R}_{\alpha \beta} \otimes {\bf R}_{\alpha \beta} \Delta V_{\beta}
\label{eq:PRD2}
\end{equation}
and it is called as the moment function, which is a second order tensor.
Note that in Eqs. (\ref{eq:PRD1}) and (\ref{eq:PRD2}),
$\omega (|{\bf R}_{\alpha i}|)$ is a localized window function, and
the common choices are the Gaussian function or the cubic spline function.

The Gaussian is defined as
\begin{equation}
\omega_h ({\bf x}) = {1 \over (\pi h^2)^{d/2} } \exp
\Bigl(
- { {\bf x} \cdot {\bf x} \over h^2 }
\Bigr)
\end{equation}
The following cubic spline function is also often
chosen in the computation,
\begin{equation}
\omega_h (q) = { A \over h^d}
\left \{
\begin{array}{lcl}
\displaystyle
1 - {3 \over 2} q^2 + {3 \over 4} q^3, &{}&~0\le q <1
\\
\\
\displaystyle
{1 \over 4} (2-q)^3, &{}& ~1\le q \le 2
\\
\\
\displaystyle
0, &{}&~{\rm otherwise}
\end{array}
\right .
\end{equation}
where $d$ is number of space dimension, $h$ is the support size,
and
\[
A = \left \{
\begin{array}{lcl}
{2 / 3} &{}&~1d
\\
{10 / (2 \pi)} &{}&~2d
\\
{1 / \pi} &{}&~3d
\end{array}
\right .
\]

If we assume that the Cauchy-Born rule may be applied
for the coarse scale displacement field, i.e.
\begin{equation}
{\bf r}_{\alpha \beta} =
{\bf F}_{\alpha} {\bf R}_{\alpha \beta}~.
\label{eq:Coarse-grain-CB}
\end{equation}
By substituting (\ref{eq:Coarse-grain-CB}) into
(\ref{eq:PRD1}), we can obtain,
\begin{eqnarray}
{\bf F}_{\alpha} &=& \Bigl( \sum_{\beta=1}^{N_h}
\omega (|{\bf R}_{\alpha \beta}|)
{\bf r}_{\alpha \beta} \otimes {\bf R}_{\alpha \beta} \Delta V_{\beta}
\Bigr) \cdot {\bf K}^{-1}_{\alpha}
\nonumber
\\
&=&
\Bigl( \sum_{\beta=1}^{N_h}
\omega (|{\bf R}_{\alpha \beta}|)
{\bf F}_{\alpha} {\bf R}_{\alpha \beta}
\otimes {\bf R}_{\alpha \beta} \Delta V_{\beta}
\Bigr) \cdot {\bf K}^{-1}_{\alpha}
\nonumber
\\
&=& {\bf F}_{\alpha}
\nonumber
\end{eqnarray}
\begin{figure}
\begin{center}
\includegraphics[width=4.0in]{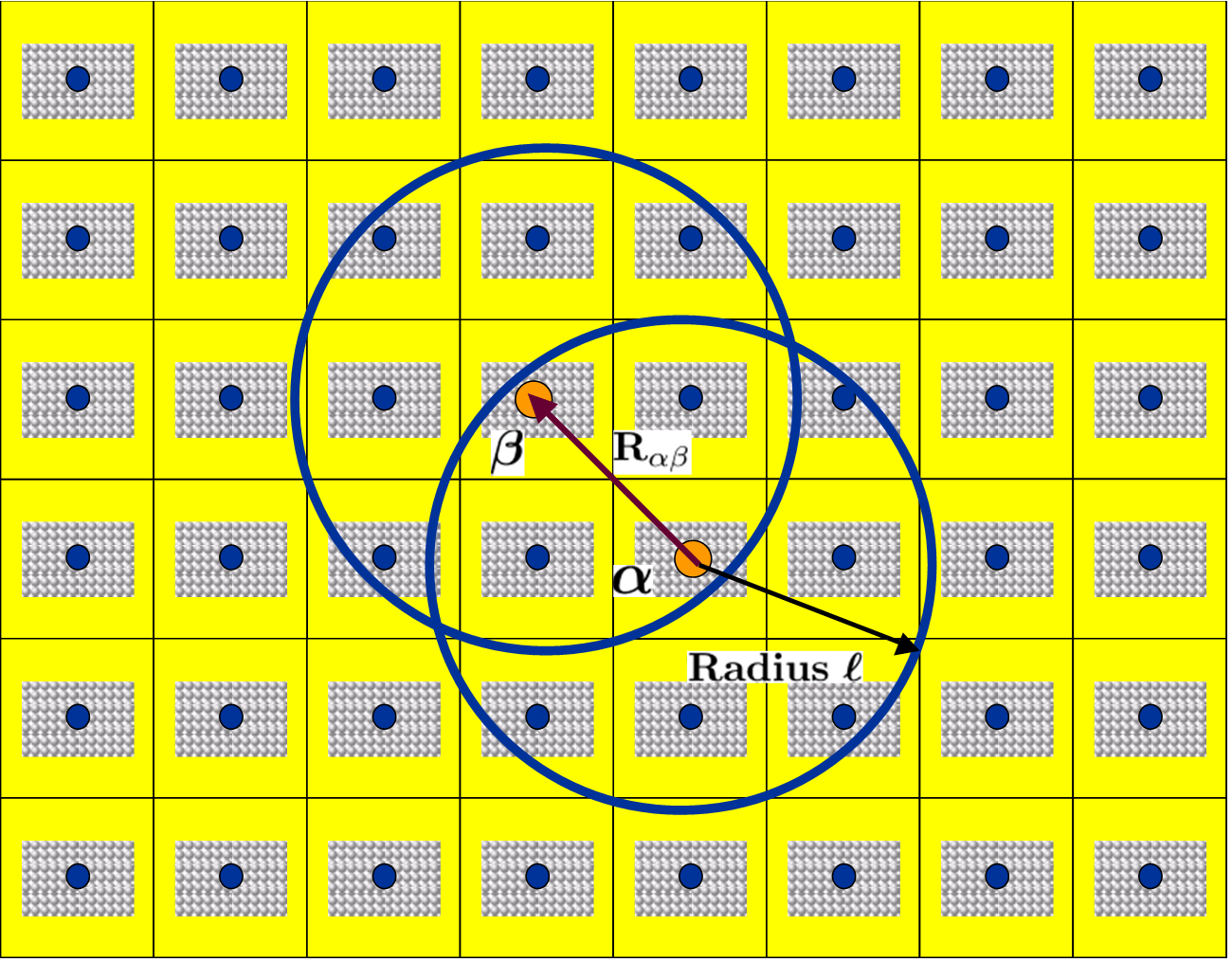}
\end{center}
\caption{Nonlocal multiscale micromorphic particle dynamics}
\label{fig:Peri-MD}
\end{figure}
We can use all the formulas that are derived in the last Section
without any modification, except that we need to explicitly
evaluate,
\begin{equation}
{\partial {\bf F}_{\alpha} \over \partial {\bf r}_{\alpha}}
= - \Bigl( \sum_{\beta=1}^{N_h}
\omega (|{\bf R}_{\alpha \beta}|)
{\bf I}^{(2)} \otimes {\bf R}_{\alpha \beta} \Delta V_{\beta}
\Bigr) \cdot {\bf K}^{-1}_{\alpha} =
- {\bf I}^{(2)} \otimes {\bf D}_{\alpha}, ~~\beta \not = \alpha,
\end{equation}
where
\[
{\bf D}_{\alpha} =
\Bigl( \sum_{\beta=1}^{N_h}
\omega (|{\bf R}_{\alpha \beta}|)
{\bf R}_{\alpha \beta} \Delta V_{\beta}
\Bigr) \cdot \mathcal{K}^{-1}_{\alpha}~
\]
is a vector.

The same is true for the time derivative of deformation gradient, i.e.
\begin{equation}
{\partial \dot{\bf F}_{\alpha} \over \partial \dot{\bf r}_{\alpha}}
= - \Bigl( \sum_{\beta=1}^{N_h}
\omega (|{\bf R}_{\alpha \beta}|)
{\bf I}^{(2)} \otimes {\bf R}_{\alpha \beta} \Delta V_{\beta}
\Bigr) \cdot {\bf K}^{-1}_{\alpha} =
- {\bf I}^{(2)} \otimes {\bf D}_{\alpha},~~\beta \not = \alpha.
\end{equation}
In general,
\begin{equation}
{\partial {\bf F}_{\beta} \over \partial {\bf r}_{\alpha}}
= \Bigl( \sum_{\gamma=1}^{N_h}
\omega (|{\bf R}_{\beta \gamma}|)
(\delta_{\alpha \gamma} - \delta_{\alpha \beta} ){\bf I}^{(2)}
\otimes {\bf R}_{\beta \gamma} \Delta V_{\gamma}
\Bigr) \cdot {\bf K}^{-1}_{\beta} =
- {\bf I}^{(2)} \otimes {\bf D}_{\beta},~~~\gamma \not = \beta~,
\end{equation}
and
\begin{equation}
{\partial \dot{\bf F}_{\beta} \over \partial \dot{\bf r}_{\alpha}}
= \Bigl( \sum_{\gamma=1}^{N_h}
\omega (|{\bf R}_{\beta \gamma}|)
(\delta_{\alpha \gamma} - \delta_{\alpha \beta} ){\bf I}^{(2)}
\otimes {\bf R}_{\beta \gamma} \Delta V_{\gamma}
\Bigr) \cdot {\bf K}^{-1}_{\beta} =
- {\bf I}^{(2)} \otimes {\bf D}_{\beta},~~~\gamma \not = \beta~.
\end{equation}

To formula a single scale coarse grained MD, we first turn off the fine scale variables,
\[
{\bfg \chi}_{\alpha} \approx {\bf I}^{(2)}, ~\alpha = 1, 2 ,\cdots M
\]
and for each cell,
\[
{\bf S}_i \approx {\bf S}_i(0) = {\bf R}_{\alpha i}~,~~i=1,2, \cdots N_{\alpha}
\]
We then obtain a coarse scale molecular dynamics,
\begin{equation}
M_{\alpha} \ddot{\bf r}_{\alpha} + \sum_{\beta \not = \alpha}
\sum_{i \in \alpha, j \in \beta}
V^{\prime} (r_{ij}) { {\bf r}_{ij} \over |{\bf r}_{ij}|} + \mathcal{B}_{\alpha} = 0~,
\label{cgmd1}
\end{equation}
where
\begin{equation}
{\bf r}_{ij} = {\bf r}_{\alpha \beta} + {\bf F}_{\beta} \cdot {\bf R}_{\beta j}
- {\bf F}_{\alpha} \cdot {\bf R}_{\alpha i}~.
\label{cgmd2}
\end{equation}
The coarse grain dynamic equations (\ref{cgmd1}) and (\ref{cgmd2}) form a close system.

To formula a two-scale coarse grained MD, we only turn off the fine scale oscillation,
\[
{\bf S}_i \approx {\bf S}_i(0) = {\bf R}_{\alpha i}~,~~i=1,2, \cdots N_{\alpha}~.
\]
The governing equations of the two-scale molecular dynamics are,
\begin{equation}
M_{\alpha} \ddot{\bf r}_{\alpha} + \sum_{\beta \not = \alpha}
\sum_{i \in \alpha, j \in \beta}
V^{\prime} (r_{ij}) { {\bf r}_{ij} \over |{\bf r}_{ij}|} + \mathcal{B}_{\alpha} = 0~,
\label{cgmd3}
\end{equation}
where
\begin{equation}
{\bf r}_{ij} = {\bf r}_{\alpha \beta} + {\bfg \phi}_{\beta} \cdot {\bf R}_{\beta j}
- {\bfg \phi}_{\alpha} \cdot {\bf R}_{\alpha i}~,~~{\rm where}~
{\bfg \phi}_{\alpha} = {\bfg \chi}_{\alpha} \cdot {\bf F}_{\alpha}~.
\label{cgmd4}
\end{equation}
The dynamic equation for micro-deformation tensor in each cell is determined by
\begin{equation}
\ddot{\bfg \phi}_{\alpha} {\bf J}_{\alpha}
+ {\bfg \phi}_{\alpha} (\mathcal{S}^{int} _{\alpha}
- \mathcal{S}^{ext}_{\alpha}) \Omega_{\alpha 0} + {\bf M}_{\alpha} = 0~,
\label{mmm1}
\end{equation}
with
\begin{eqnarray}
\mathcal{S}^{int}_{\alpha} &=& {1 \over \Omega_{\alpha 0}} \sum_{i \in \alpha}
\sum_{j \in \alpha, j \not \in i}
{ V^{\prime} (r_{ij}) \over r_{ij}}
{\bf S}_i \otimes {\bf S}_i
\label{mmm2}
\\
\mathcal{S}^{ext}_{\alpha} &=&
{1 \over \Omega_{\alpha 0}} \sum_{\beta \not = \alpha}
\sum_{i \in \alpha, j \in \beta}
{ V^{\prime} (r_{ij}) \over r_{ij}}
{\bf r}_{ij} \otimes {\bf S}_i
\label{mmm3}
\end{eqnarray}
and ${\bf M}_{\alpha} = \sum_{i \in \alpha} {\bf f}_i \otimes {\bf R}_{\alpha i} $.
Eqs. (82)-(86) are also a closed system.

\section{Discussions}
In this work,
we have proposed in the first time a novel concept
of multiplicative multiscale decomposition.
By analyzing the structure of
the (Andersen)-Parinello-Rahman molecular dynamics,
we have extended the (Andersen)-Parrinello-Rahman MD to
form a novel multiscale micromorphic molecular dynamics (MMMD)
that can solve finite size molecular dynamics
problems without the restriction of the periodic boundary condition.
In other words, it can solve finite size molecular dynamics problems
with arbitrary boundary condition.
This is because we can apply boundary conditions to the coarse
scale variables, say ${\bf r}_{\alpha}$, to impose
the boundary conditions at macroscale.

Different from the most multiscale methods proposed in recent years,
the proposed multiscale dynamics formulation is not aimed
at saving computation time or resource,
but aimed at revealing multiscale connections and structures
so that we can apply molecular dynamics
to solve engineering problems with arbitrarily domain
and general boundary condition.
It is the author's opinion that if only we can achieve these goals
we can start to think about how to build a coarse-grain model that can provide
the efficient computing and save computational resources.

The conventional wisdom is that
if we simply increase the size of molecular dynamics
simulation we can simulate large and large size of objects
based on the first principle.
In order to capture correct thermodynamics response of a finite
size molecular system,
we cannot only solve massive numbers Newton equations,
instead the system's multiscale characters must be
carefully taken into account so that the microscale quantities
can be correctly related to mesoacel and macroscale quantities based
on first principle. It may be noted that the multiscale
technique employed here is not for saving computer resource
but for correct simulations of thermodynamic variables for
a finite size system.
Moreover, the MMMD formulation is essentially a local
N$\mathcal{S}$H ensemble formulation, and we have not considered
the thermal or temperature effects yet.
A future study to extend the present theoretical formulation to
other molecular dynamics ensembles such as local N${\bfg \phi}$T ensemble 
will be reported in a separated paper, and
the computer implementation of the multiscale micromorphic
molecular dynamics formulation will be reported in the second part
of this work.
%%%%

\begin{acknowledgments}
The author would like to thank Professor Antonio DiCarlo
for the enlightening discussions on the subject.
Q. Tong is supported by a graduate fellowship from Chinese Scholar
Council (CSC), and this support is greatly appreciated.
\end{acknowledgments}

\bibliographystyle{jacs-new}
\bibliography{prmd}

\begin{thebibliography}{10}

\bibitem{Andersen80}
Andersen, H.~C.
\newblock {\em Journal of Chemical Physics}, {\bf 1980}, {\em 72}, 2384--2393.

\bibitem{PR80}
Parrinello, M.; Rahman, A.
\newblock {\em Physical Review Letters}, {\bf 1980}, {\em 14}, 1196--1199.

\bibitem{PR81}
Parrinello, M.; Rahman, A.
\newblock {\em Journal of Applied Physics}, {\bf 1981}, {\em 12}, 7182--7190.

\bibitem{Parrinello02}
Laio, A.; Parrinello, M.
\newblock {\em Proceedings of National Academy of Science, USA}, {\bf 2002},
  {\em 99}, 12562--12566.

\bibitem{Parrinello03}
Marton\'{a}k, R.; Laio, A.; Parrinello, M.
\newblock {\em Physical Review Letters}, {\bf 2003}, {\em 90}, 075503.

\bibitem{Parrinello06}
Marton\'{a}k, R.; Donadio, D.; Oganov, A.; Parrinello, M.
\newblock {\em Nature materials}, {\bf 2006}, {\em 5}, 623--626.

\bibitem{Podio08}
Podio-Guidugli, P.
\newblock {\em Journal of Elasticity}, {\bf 2010}, {\em 100}, 145--153.

\bibitem{Rudd-Broughton1}
Rudd, R.~E.; Broughton, J.~Q.
\newblock {\em Physical Review B}, {\bf 1998}, {\em 58}, R5893--R5896.

\bibitem{Rudd-Broughton2}
Rudd, R.~E.; Broughton, J.~Q.
\newblock {\em Physical Review B}, {\bf 2005}, {\em 72}, 144104.

\bibitem{Youping03}
Chen, Y.; Lee, J.~D.
\newblock {\em Physica A}, {\bf 2003}, {\em 322}, 359--376.

\bibitem{Youping09}
Chen, Y.
\newblock {\em The Journal of Chemical Physics}, {\bf 2009}, {\em 130}, 134706.

\bibitem{Voth1}
Noid, W.; Chu, J.-W.; Ayton, G.~S.; Krishna, V.; Izvekov, S.; andA. Das, G.
  V.~V.; Andersen, H.~C.
\newblock {\em The Journal of Chemical Physics}, {\bf 2008}, {\em 128}, 244114.

\bibitem{Voth2}
Noid, W.; Liu, P.; Wang, Y.; Chu, J.-W.; Ayton, G.~S.; Krishna, V.; Izvekov,
  S.; Andersen, H.~C.; Voth, G.~V.
\newblock {\em The Journal of Chemical Physics}, {\bf 2008}, {\em 128}, 244115.

\bibitem{DiCarlo09}
DiCarlo, A.
\newblock {\em Private Communications}, {\bf 2009}, pp 1--2.

\bibitem{Ray84}
Ray, J.~R.; A.Rahman.
\newblock {\em Journal of Chemical Physics}, {\bf 1984}, {\em 80}, 4423--4428.

\bibitem{Marsden-Hughes}
Marsden, J.; Hughes, T.
\newblock {\em Mathematical Foundations of Elasticity}.
\newblock Prentice-Hall, Inc., 1983.

\bibitem{Irving50}
Irving, J.; Kirkwood, J.~G.
\newblock {\em The Journal of Chemical Physics}, {\bf 1950}, {\em 18},
  817--829.

\bibitem{Tsai79}
Tsai, D.~H.
\newblock {\em The Journal of Chemical Physics}, {\bf 1979}, {\em 70}, 1375.

\bibitem{Zhou03}
Zhou, M.
\newblock {\em Proceedings of The Royal Society of London. Series A}, {\bf
  2003}, {\em 459}, 2347--2392.

\bibitem{Murdoch07}
Murdoch, A.
\newblock {\em Journal of Elasticity}, {\bf 2007}, {\em 88}, 113--140.

\bibitem{Sun08}
Subramaniyan, A.~K.; Sun, C.~T.
\newblock {\em International Journal of Solids and Structures}, {\bf 2008},
  {\em 45}, 4340--4346.

\bibitem{Li-Liu99}
Li, S.; Liu, W.
\newblock {\em International Journal of Numerical Methods for Engineering},
  {\bf 1999}, {\em 45}, 251.

\bibitem{Silling2}
Silling, S.; Epton, M.; Weckner, O.; Xu, J.; Askari, E.
\newblock {\em Journal of Elasticity}, {\bf 2007}, {\em 88}, 151--184.

\end{thebibliography}

\end{document}